\documentclass{article}

\usepackage[final]{neurips_2019}

\usepackage[utf8]{inputenc} 
\usepackage[T1]{fontenc}    
\usepackage{hyperref}       
\usepackage{url}            
\usepackage{booktabs}       
\usepackage{amsfonts}       
\usepackage{nicefrac}       
\usepackage{microtype}      
\usepackage{graphicx}  
\newtheorem{theorem}{Theorem}

\title{Reinforcement Learning for Market Making in a Multi-agent Dealer Market}

\author{%
    Sumitra Ganesh  \\
  JPMorgan AI Research\\
  \And
  Nelson Vadori\\
  JPMorgan AI Research\\
  \And
  Mengda Xu \\
  JPMorgan AI Research \\
  \And
  Hua Zheng \\
  JPMorgan Quantitative Research \\
  \And
  Prashant Reddy \\
  JPMorgan AI Research\\
  \And
  Manuela Veloso \\
  JPMorgan AI Research\\
}

\begin{document}

\maketitle

\begin{abstract}
Market makers play an important role in providing liquidity to markets by continuously quoting prices at which they are willing to buy and sell, and managing inventory risk.  In this paper, we build a multi-agent simulation of a dealer market and demonstrate that it can be used to understand the behavior of a reinforcement learning (RL) based market maker agent. We use the simulator to train an RL-based market maker agent with different competitive scenarios, reward formulations and market price trends (drifts). We show that the reinforcement learning agent is able to learn about its competitor's pricing policy; it also learns to manage inventory by smartly selecting asymmetric prices on the buy and sell sides (skewing), and maintaining a positive (or negative) inventory depending on whether the market price drift is positive (or negative). Finally, we propose and test reward formulations for creating risk averse RL-based market maker agents. 
\end{abstract}

\section{Introduction}
\label{intro}
Deep reinforcement learning (RL) algorithms have had considerable success in surpassing human level performance in several single and multi-agent learning problems. A key ingredient of these advances has been the availability of gaming and physics simulators which provide a ready stream of experience for developing, training and testing algorithms. However, the financial domain lacks a similar suite of simulators that could be used to conduct research. 

Simulators provide more than just data for sample-intensive RL algorithms; they also provide a platform to conduct controlled experiments to test what is being "learnt" by an agent, how a policy performs in different scenarios and the causality between changes in the environment and agent behavior. Simulators can also be used to train an agent in a diverse set of scenarios, leading to improved generalization and robustness to changes in the environment.  In this paper we build a multi-agent simulator for a \textit{dealer market} and demonstrate that it can be used to train an RL-based \textit{market maker} agent and understand its behavior. 

Market makers play an important role in providing liquidity to markets by continuously quoting prices at which they are willing to buy and sell. 
In this paper, we focus on a dealer market (also commonly known as an over-the-counter or OTC market) where there is a single security being traded and market makers continuously stream prices to buyers and sellers (referred to as \textit{investors}). The investors, who come to the market to execute their orders, observe these prices and select the market maker they want to trade with. The two types of agents - market makers and investors - interact directly with each other, and only observe the transactions that they are involved in and reference price information from an exchange. 

A marker maker's policies around pricing and risk management in this market depend on its objectives and preferences (e.g. how risk averse it is), the policies of competing market makers, the overall market environment (e.g. volatility) and trade flow from investors. Our objective in this paper is to use the multi-agent simulator to study the performance of a reinforcement learning (RL) based market making agent under different competitive scenarios, reward formulations and market environments. 

In order to model different competitive scenarios, we formalize the dealer market as a multi-agent system and build a simulator with the ability to instantiate market maker agents with different policies. In particular, we implement an adaptive market maker agent which acts as a realistic benchmark and competitor for the RL agent. We then train an RL-based market maker against different competitors, with varying reward formulations and market environment conditions, and study its behavior and performance. 

\subsection{Related Work}
The market making problem has been extensively studied in finance and economics literature, largely as an optimal control problem. Classical models such as \cite{Garman1976-oy}, \cite{Amihud1980-qz}, \cite{Ho1981-yy}, and more recently \cite{Avellaneda2008-de} and \cite{Gueant2013-gc}, focus on the role of inventory risk in determining the optimal pricing strategy for a market maker. Other models, such as \cite{Glosten1985-cx}, study the role of adverse selection risk arising from informed traders in the market. All these works model a single market maker and make assumptions about the distributions of order arrivals and price to derive analytical solutions for the market maker's pricing policy using stochastic optimal control techniques. In contrast, in our approach the distribution of order arrivals for a market maker naturally arises from the strategies and attributes of other agents in the system.

Market making has also been studied in the agent-based modeling (ABM) literature (e.g. \cite{Darley2000-aw}, \cite{Das2005-ji},\cite{Das2008-mu}, \cite{Jumadinova2010-pu}, \cite{Wah2017-lb}), but the agents tend to be simplistic with a view to studying market equilibria and endogeneous price formation rather than developing realistic market making strategies. 

Most previous work focuses on market making in \textit{limit order book} markets where all agents submit their orders to a central matching facility and can observe all outstanding orders and executed transactions in the market. In contrast, there has been relatively little work focused on dealer markets (\cite{Gueant2017-cw}, \cite{Bank2018-ip}, \cite{Ghoshal2016-kj})where the agents interact directly with each other and only observe the trades they are involved in. Our multi-agent simulation realistically replicates the partial observability conditions of a dealer market allowing us to study the challenges it poses for reinforcement learning algorithms. 

\cite{Chan2001-hp}, and more recently, \cite{Lim2018-yh} and \cite{Spooner2018-vj} have developed RL-based market making approaches for limit order book markets; however, they do not explicitly model the competing market makers or study different competitive scenarios.

\subsection{Our Contributions}

\noindent We make the following contributions in this paper: \textbf{(i)} we formalize the dealer market as a multi-agent system (section  \ref{sec:multi-agent-sim}); \textbf{(ii)} we define an adaptive market making policy inspired by the mean-variance optimization techniques commonly used in finance (section \ref{sec:adaptive-agent}).; \textbf{(iii)} we conduct extensive experiments with competing market maker agents and are able to draw some key insights about what the RL-based market maker is learning. (sections \ref{sec:rl-agent}, \ref{sec:experiments}). In particular, we show that the RL-based market maker is able to learn about its competitor's pricing strategy and manage inventory by skewing prices; it also learns to exploit any significant market price drift by building inventory; \textbf{(iv)} we propose and test several reward formulations that can be used to create a risk-averse RL-based market maker (section \ref{rl-reward-exp}).

We start by defining the mechanics of the dealer market and key terminology in the next section.



\section{Dealer Market - Definitions}
\label{sec:dealer-markets}

\noindent \textbf{Prices and Spreads}: In a dealer market with a single traded security, each market maker has to stream two prices at each time step - the price at which it is willing to buy (\textit{bid}) and the price at which it is willing to sell (\textit{ask}). We assume that there is also an exchange limit order book market (LOB) for the same security, and that all market makers receive reference pricing information from this exchange. 

The exchange reference price can be broken down into two components: (a) the \textit{mid-price} $P$, which is defined as the average of the best bid and ask in the LOB, and (b) the \textit{spread} $S^{b}_{\mathrm{ref}}$ or $S^{s}_{\mathrm{ref}}$, which is defined as the difference between the price of a buy/sell trade on the exchange and the mid-price. By convention, market makers also quote their buy/sell prices relative to the exchange mid-price i.e. the $i$-th market maker quotes spreads $S^{b}_i$ (buy) and $S^{s}_i$ (sell). Note that this implies that the market maker buys at $P_t - S^{b}_i$ and sells at $P_t + S^{s}_i$. 

\noindent \textbf{Spread PnL}: For each trade with an investor, the market maker earns a profit that is referred to as \textit{spread PnL}. In practice, the spreads are a function of the trade size. Thus, for a trade of size $v$, the market maker earns a profit of $v \cdot S^b_i(v)$ if it is a buy, and $v \cdot S^s_i(v)$ if it is a sell. 

\noindent \textbf{Inventory}: As the market maker executes trades with investors, it will typically accumulate an \textit{inventory} $z_t$, which could be positive or negative depending on whether the agent has bought more than it sold up to time $t$, or vice versa. By convention, buys are positive and sells are negative in the inventory calculation. \footnote{Buys and sells are from the market maker's perspective unless otherwise noted.}

\noindent \textbf{Inventory PnL}: A non-zero inventory exposes the market maker to risk from mid-price fluctuations.  If the market maker holds an inventory of $z_t$ from $t$ to $t+1$, and the mid-price changes from $P_t$ to $P_{t+1}$, the value of the inventory changes by $(P_{t+1}-P_t)\cdot z_t$. 

\noindent \textbf{Hedging}: Ideally, a market maker would like to keep a small inventory so as to minimize risk. The market maker could choose to reduce risk by initiating a trade which reduces its inventory (e.g. try to buy if it has negative inventory) - this is referred to as \textit{hedging}. In doing so, it incurs a cost since it assumes the role of a price-taker in the hedging transaction. 


\noindent \textbf{Internalization and Skewing}: The market maker could also try to reduce risk organically by \textit{skewing} its prices to attract trades that offset its inventory. For example, if it has a positive inventory, it could make its buy prices less attractive and sell prices more attractive to investors. The effect of organic risk mitigation by offsetting investor trades is referred to as \textit{internalizaton}.

\noindent \textbf{Market share}: As multiple market makers compete in a dealer market, each one captures a portion of the investor trade flow depending on its pricing. The percentage of total trades captured by a market maker, weighted by trade size, is referred to as its market share.



\section{Multi-Agent Simulator}
\label{sec:multi-agent-sim}
We consider a dealer market with $M$ market makers and $N$ investors and a single traded security. A trade between market maker $i$ and investor $j$ at time $t$ is denoted by $v_{i,j,t}$; buys are positive, sells are negative and zero implies no trade.

\subsection{Market Maker Agent}
\label{mm-agent-def}
The observations, actions and rewards for the $i$-th market maker at time $t$ are summarized below. The market maker receives observations about investor trades it won in the previous time step and reference prices for current time step. It then decides on the prices to stream for time step $t$ and any hedge trades to execute. Finally, it receives a set of rewards in the form of spread PnL, hedge cost and inventory PnL. 

\noindent \textbf{Observations.} (i) Trades executed in previous time step: $\{v_{i,j,t-1}\}_{j=1:N}$; (ii) Inventory $z_{i,t} = z_{i,t-1} + \sum_{j=1:N} v_{i, j, t-1}$; (iii) Reference mid-price $P_{t}$ and spread curves $S^b_{ref, t} (v)$ and $S^s_{ref, t} (v)$; (iv) Market share, or equivalently, total market volume in previous time step.

\noindent \textbf{Actions.} (i) Pricing: Determine spreads $S^b_{i, t} (v)$ and $S^s_{i, t} (v)$ to stream; (ii) Hedging: Determine fraction $x_{i,t}$ of the inventory $z_{i,t}$ to hedge.

\noindent \textbf{Rewards.} (i) Spread PnL: For each investor trade $v$, the market maker earns $|v| \cdot S^b_{i, t}(v)$ or $|v| \cdot S^s_{i, t}(v)$, depending on whether it is a buy/sell; (ii) Hedge Cost: Market maker pays a cost of $|v| \cdot S^b_{ref, t}(v)$ if it is selling and $|v|\cdot S^s_{ref, t}(v)$ if it is buying, where $v = - x_{i,t} z_{i,t}$; assuming that the market maker hedges with the exchange; (iii) Inventory PnL: $(P_{t+1}-P_t)\cdot z_t$; (iv) Total PnL =Sspread PnL + Hedge Cost + Inventory PnL.

  
\noindent In this paper, we assume that the buy spread streamed by a market maker agent is of the form $ S^b_{t}(v) := S^b_{ref, t}(v) \cdot ( 1 + \epsilon_b)$, where the market maker subscript is dropped for simplicity (a similar relation is assumed for the sell spread). Thus, the pricing action reduces to a choice of $\epsilon_b$ and $\epsilon_s$ at each time step $t$. 

Based on this parametrization, we can define two simple baseline market maker agents: (a) \textbf{Random Agent} that picks $\epsilon_b, \epsilon_s \sim \mathrm{Unif}[\epsilon_{min}, \epsilon_{max}]$, and the fraction to hedge $x \sim \mathrm{Unif}[0,1]$ at each time step (b) \textbf{Persistent Agent} that keeps a fixed $\epsilon_b, \epsilon_s$ and hedges a fixed fraction $x$ at each time step.


\subsection{Investor Agent}
We assume that each investor agent has a trade generation process specified by probability distributions for trade size, direction and arrival.  The investor agent executes the trade in the same time step by picking the market maker to trade with that offers him the best price.






\subsection{Exchange reference price}
\label{srefmidprice}
The mid price process $P$ is diffused as a geometric brownian motion with drift $\mu$ and volatility $\sigma$, i.e. $P_t=P_{t-1} \exp(\mu \Delta - 0.5 \sigma^2\Delta +\sigma \sqrt{\Delta}Z_t)$, where $(Z_t)$ are i.i.d. standard normal and $\Delta$ is the timestep value ($\Delta=$15 mins). Note that $E[P_t]=P_0\exp(\mu \Delta)$.

The reference spread curves $S^s_{ref, t} (v)$, $S^b_{ref, t} (v)$ are assumed to be equal and are sampled at each timestep from a statistical model calibrated to real market data (details in Appendix).

\section{Adaptive Market Maker Agent}
\label{sec:adaptive-agent}

In this section, we will define a more realistic, adaptive pricing and hedging policy for the market maker agent. The pricing/hedging algorithms used in this section are inspired by standard mean-variance trade-off formulations used in the mathematical finance literature for pricing and hedging, cf \cite{AC}, \cite{Bank2018-ip}, further tailored to suit our framework.

We assume that the adaptive market maker maintains an \textit{empirical response table} with estimates of the mean and variances of the incoming net flow $v_\epsilon$ and the normalized spread PnL $s_\epsilon$ (normalized by $S_{\mathrm{ref}}(0)$) as a function of $(\epsilon_b, \epsilon_s)$. In our experiments, the agent uses its observations at each time step to update the table with an exponential forgetting factor of $\beta$ (set to 0.35 ). We also assume that the agent holds an estimate of the mid-price volatility $\sigma$, below we will use the normal volatility $\hat{\sigma}:= \sigma \sqrt{\Delta} P_{t}$ over the each interval $[t,t+1]$. 

Overall, the adaptive market maker's policy is determined by two static parameters - market share target $\eta_{ms}$ and risk aversion $\gamma$, and the empirical response table which is dynamically updated based on the agent's observations. 
Results showing that the adaptive agent outperforms simple baseline agents are included in the Appendix. 
\subsection{Adaptive Pricing Policy}
\label{sec:adaptive-pricing-policy}

The adaptive market maker chooses $(\epsilon_b, \epsilon_s)$ at each time step to (a) meet its market share target $\eta_{ms}$, and (b) internalize by skewing its prices. It does this in two steps:

\noindent \textbf{Step 1}: The market maker picks $\epsilon_b = \epsilon_s = \epsilon_*$ to achieve its market share target $\eta_{ms}$ within a tolerance $\delta_{tol}=5\%$ by solving
$$
\epsilon_*=\max \left\{ \epsilon: \hspace{1mm} |cost_1(\epsilon)-min_x cost_1(x)| \leq \delta_{tol}\right\}; \hspace{3mm}
cost_1(\epsilon):=\left|\eta_{ms} - \frac{E[v_\epsilon]}{V_{market}}\right|
$$
\noindent where $V_{market}$ is the observed total market volume and $E[v_\epsilon]$ is looked up from the empirical estimates. This step ensures that the market maker is able to adapt its pricing if it is getting little or no trade flow. 

\noindent \textbf{Step 2}: The market maker then skews only one side (buy/sell) to attract a flow that offsets its current inventory $z$ i.e. if it has negative (positive) inventory it decreases $\epsilon_b$ ($\epsilon_s$) to attract offsetting buys (sells). However, decreasing spreads impacts spread PnL; so the market maker has to trade-off between the loss of spread PnL and the risk of carrying inventory in deciding how much to skew. It selects the optimal $\epsilon$ for the side it is skewing (while the other side is kept fixed to $\epsilon_*$) so as to minimize the below cost function:
$$
cost_2(\epsilon):= -\underbrace{S_{\mathrm{ref}}(0)E[s_\epsilon]}_{\mbox{spread PnL}}
+\gamma \sqrt{\underbrace{S_{\mathrm{ref}}(0)^2 var[s_\epsilon]}_{\mbox{spread PnL uncertainty}} + \underbrace{\hat{\sigma}^2 E[(z+v_\epsilon)^2]}_{\mbox{cost of risk + internalization}}}
$$
In the above, the mid-price increment over one timestep is assumed to be normally distributed with mean 0 and variance $\hat{\sigma}^2$; $E[s_\epsilon]$, $var[s_\epsilon]$, $E[v_\epsilon]$, $var[v_\epsilon]$ are empirical estimates; and $\gamma$ is the risk-aversion parameter representing the mean-variance trade-off.

\subsection{Adaptive Hedging Policy}
\label{sec:adaptive-hedging-policy}

The approach we choose for hedging is similar to the mean-variance approach in step 2 of \ref{sec:adaptive-pricing-policy}. In hedging, the market maker has to decide the fraction $x$ of the current inventory $z$ to hedge by initiating an offsetting trade. In doing so, the market maker has to trade off between (a) the cost of hedging (which increases with the size of the hedge trade), and (b) the risk of carrying inventory and being impacted by mid-price changes. 

The market makers's cost over one timestep is $c(x) :=|x z|S_{\mathrm{ref}}(x z) - \hat{\sigma} (z(1-x)+v_\epsilon) Z$, where $Z \sim N(0,1)$. This leads to setting $x$ so as to minimize the below cost function:
$$
cost_{hedge}(x):= \underbrace{|xz|S_{\mathrm{ref}}(x z)}_{\mbox{hedge cost}}
+\gamma \sqrt{\underbrace{\hat{\sigma}^2 E[(z(1-x)+v_\epsilon)^2]}_{\mbox{cost of risk + internalization}}}
$$
In the above, $E[v_\epsilon]$ and $var[v_\epsilon]$ are empirical estimates, and the risk aversion parameter $\gamma$ is chosen to be the same as the one using in pricing, for consistency. 

\section{RL-based Market Maker Agent}
\label{sec:rl-agent}

We use a standard implementation of the Proximal Policy Optimization (PPO) algorithm with clipped objective  in Rllib (\cite{Schulman2017-fq}, \cite{Liang2017-dn}) to train the RL-based market maker. The input to the policy network includes all the observations available to the agent at that time step (see \ref{mm-agent-def}), as well as the inventory PnL from the previous time step. The outputs of the policy network are the pricing parameters $\epsilon_b, \epsilon_s \in [-1,1]$ and the fraction to hedge  $x \in [0,1]$. By default, the RL agent uses the Total Pnl specified in \ref{mm-agent-def} as the scalar reward to train its policy. 

\subsection{Optimal pricing for spread PnL}
Our primary objective is to understand the behavior of an RL-based market maker when it is trained against different competitors. In order to assess whether the RL agent is learning about its competitor's policy, we derive the optimal $\epsilon_b, \epsilon_s$ in Theorem \ref{thmopteps} for a simplified case where the RL agent competes with exactly one other market maker. (proof in Appendix).

\begin{theorem} 
\label{thmopteps}
The expected spread PnL of an agent competing with a market maker agent that quotes buy/sell prices according to the distributions $F^{comp}_{j}(x) := P[\epsilon^{comp}_j \leq x]$, $j \in \{b,s\}$, is maximized by setting:
$$
\epsilon_j= \mbox{argmax}_\epsilon (1+\epsilon) (1-F^{comp}_{j}(\epsilon)
+p \hat{F}^{comp}_{j}(\epsilon))
$$
where p is the probability that the agent wins the trade when both agents quote the same price, $\hat{F}^{comp}_{j}(\epsilon):=P[\epsilon^{comp}_j =\epsilon]=F^{comp}_{j}(\epsilon)-F^{comp}_{j}(\epsilon^-)$. The result holds under the assumption that investor trades are possibly random but independent from $\epsilon^{comp}_b$, $\epsilon^{comp}_s$. In particular, if $F^{comp}_{b}$, $F^{comp}_{s}$ are $\mathrm{Unif}[a, b]$, with $-1 \leq a \leq b$, then we get $\epsilon^*_b=\epsilon^*_s= \frac{1}{2}\max(2a,b-1)$.
\end{theorem}

\subsection{Reward formulations for risk aversion}
Our basic reward definition of Total PnL models an agent that is not risk averse. However, in practice, a market maker would not prefer its inventory PnL to be volatile, e.g. high losses combined with high gains. To model this risk aversion, we add a penalty term to the Total PnL which aims at reducing the inventory PnL variance. This aversion to mid-price fluctuations is modeled in the adaptive market maker via its risk aversion coefficient. We propose the following risk penalty functions and empirically compare them in our experiments.
\begin{itemize}
    \item \textbf{[InvPnL Stdev]} $\mbox{Penalty}=-\alpha \cdot Std_{10}(\mbox{InventoryPnL})$ (rolling 10-timestep Stdev).
    \item \textbf{[$\mbox{InvPnL}^2$]} $\mbox{Penalty}= - \alpha \cdot \mbox{InventoryPnL}^2$
    \item \textbf{[Asymmetric]} $\mbox{Penalty}= \alpha \cdot\min(0,\mbox{InventoryPnL})$
\end{itemize}
\noindent In our experiments $\alpha=0.4$, $0.07$ and $0.3$, respectively, for the three penalty functions. These values were chosen by trial-and-error to be ranges where the RL agent still chooses to trade; a very large penalty term results in the RL agent choosing not to trade at all by setting a high spread. 

\subsection{Experimental settings}
In our experiments, we use a 2-layer fully-connected neural network with 256 nodes in each layer and {\it tanh} activation functions for the PPO policy network and a PPO clip parameter of 0.3. For each training iteration, we take one step of minibatch SGD via Adam with a minibatch size of 256 and learning rate of 5e-5. 

For each of experiments, we trained an RL policy starting from 5 different initial random seeds. In experiments with the RL agent, the agent is first trained against the competitor until convergence; the final policy is then used to calculate performance measures.

\section{Experiments}
\label{sec:experiments}

All experiments are conducted with a set of 20 investor agents, each of which generate a buy or sell order of unit size at each time step with equal probability. The exchange mid-price has zero drift ($\mu=0$) and volatility ($\sigma=0.1$), unless otherwise specified. 

In all our experiments, we use the Total Pnl defined in \ref{mm-agent-def} as the key performance measure to compare market maker agents. For a given set of agent policies, we evaluate metrics per time-step, averaged across 5000 or more simulation timesteps and 5 random seeds. All PnL metrics are multiplied by 100 for readability.


\subsection{Competitive Scenarios}

We trained an RL-based market maker against competing random, persistent and adaptive market makers. In all experiments, the PPO policy was able to converge (convergence plots in Appendix).

\noindent \textbf{Performance}: 
The RL agent was able to outperform the random and persistent agents in all experiments (Table \ref{tab:rl-exp-perf}). It also clearly outperforms the adaptive agent when the adaptive agent is risk averse ($\gamma=2$); but the adaptive agent is able to close the gap/outperform at lower risk aversion (Table \ref{tab:rl-adaptive-perf}). 

\begin{figure}[!tbp]
\centering
\begin{minipage}[b]{0.45\linewidth}
  \begin{minipage}[b]{0.45\linewidth}
    \includegraphics[width=\columnwidth]{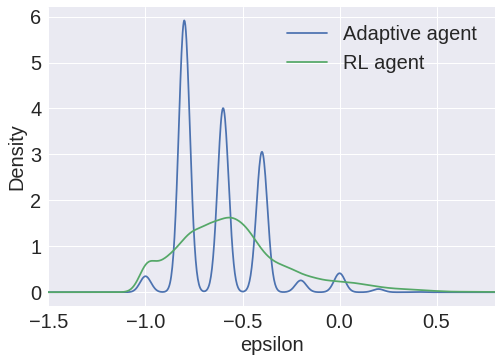}
  \end{minipage}
  \hfill
  \begin{minipage}[b]{0.45\linewidth}
    \includegraphics[width=\columnwidth]{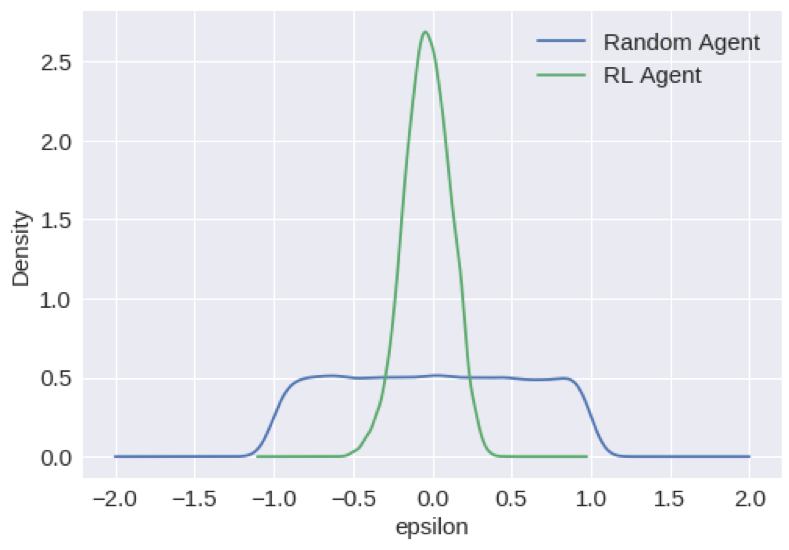}
  \end{minipage}
  \caption{RL Agent vs Competitor: $\epsilon$-distribution. The optimal pricing point according to Theorem \ref{thmopteps} is 0 against the random agent and -0.6 against the adaptive agent.}
  \label{fig:rl-eps-compare}
\end{minipage}
\hfill
\begin{minipage}[b]{0.45\linewidth}
  \begin{minipage}[b]{0.45\linewidth}
  \includegraphics[width=\columnwidth]{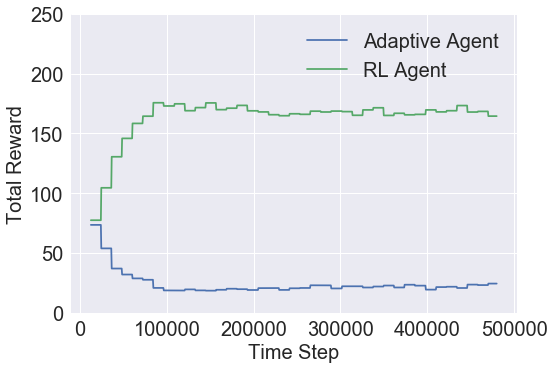}
   \end{minipage}
   \hfill
   \begin{minipage}[b]{0.45\linewidth}
   \includegraphics[width=\columnwidth]{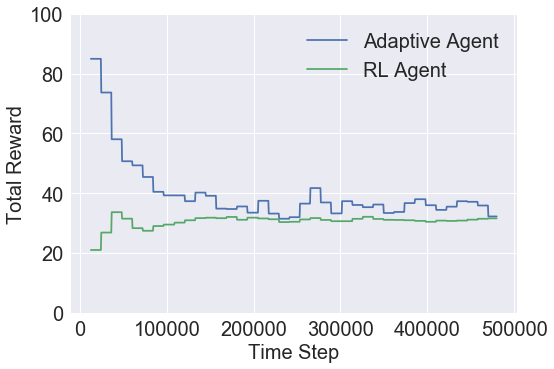}
  \end{minipage}
   \caption{RL vs Adaptive agent: Convergence during training. Left: adaptive agent - $\eta_{ms}=0.5, \gamma=2$. Right: adaptive agent - $\eta_{ms}=0.5, \gamma=0$}
   \label{fig:rl-adaptive-convergence}
\end{minipage}
\end{figure}

Figure \ref{fig:rl-adaptive-convergence} shows the convergence of the total reward for the RL and adaptive agents during training. Unlike the random or persistent agents which have fixed policies, the adaptive agent actively responds to changes in RL agent's policy by adapting its empirical response table. Note that the RL agent in this case is not risk averse and pursues Total PnL. If the adaptive agent is not risk averse  ($\gamma=0$) and targets a $0.5$ market share, it is able to converge to a similar reward as the RL agent; but it loses if it is risk averse ($\gamma=2$). 

   \begin{table}[t]
        \begin{minipage}{0.45\linewidth}
        \caption{RL Agent vs Random/Persistent: Average excess PnL per time-step of RL agent over competitor}
        \label{tab:rl-exp-perf}
         \centering
            \resizebox{\columnwidth}{!}{\smallskip 

\begin{tabular}{lrrrr}
\hline
Competitor      &  Spread PnL &  Inventory PnL &  Hedge Cost &  Total PnL \\

\hline
Persistent (-0.5) &      21.60 &         0.40 &       8.10 &     29.30 \\
Persistent (0.0)  &      74.20 &          0.20 &       1.80 &     76.20 \\
Random ($\mathrm{Unif}[-0.5,0.5]$) &      33.90 &      0.40 &      25.50 &     59.70 \\
Random ($\mathrm{Unif}[-1.0,1.0]$)    &      24.00 &         -0.70 &      30.80 &    54.10 \\
\hline
\end{tabular}

}
        \end{minipage}
        \hfill
        \begin{minipage}{0.45\linewidth}
        \centering
        \caption{RL vs Adaptive agent: Average excess PnL per time-step of RL agent over competitor}
        \label{tab:rl-adaptive-perf}
        \resizebox{\columnwidth}{!}{\smallskip 

\begin{tabular}{llrrrr}
\hline
Risk Aversion  & MS target &  Hedge Cost & Inventory PnL &  Spread PnL & Total PnL \\
\hline
 0    &  0.25 & -0.73 &           0.14 & 81.83 &      81.24 \\
      &  0.50 & -0.00 &          3.52 & -7.92 &      -4.40 \\
 1    &  0.25 &-0.37 &           0.30 & 94.16 &      94.09 \\
      &  0.50 & 0.74 &           0.01 &  4.88 &       5.64 \\
 2    &  0.25 &-0.22 &          -0.49 & 93.18 &      92.47 \\
      &  0.50 & 2.02 &           0.54 & 23.87 &      26.42 \\
\hline
\end{tabular}
}
        \end{minipage}
    \end{table}

\noindent \textbf{Learning the competitor pricing strategy}: Using Theorem \ref{thmopteps}, we can compute the optimal strategy for spread PnL maximization. In our experiments (see Table \ref{tab:rl-exp-eps}), we find that the RL agent prices, on average, just below the persistent agent as expected. Against the random agents, it prices, on average, close to the "optimal point" of 0 against $\mathrm{Unif}[-1,1]$ and -0.25 against $\mathrm{Unif}[-0.5, 0.5]$.

Furthermore, if we look at the distribution of $\epsilon_b$ for the RL agent (Figure \ref{fig:rl-eps-compare}), we find that the peak of the distribution corresponds to the "optimal pricing point", i.e. $\epsilon=0$ against the random agent and $\epsilon=-0.6$ against adaptive agent. In the experiment with the adaptive agent, we use the empirical c.d.f. of $\epsilon_b$ for the adaptive agent after convergence (in blue) and theorem \ref{thmopteps} to compute the "optimal pricing point" for the RL agent i.e. $\epsilon=-0.6$; this coincides with highest RL pricing density (green).

Note, that the "optimal pricing points" are only with respect to spread PnL, whereas the RL agent is maximizing the Total PnL. However, in practice, the average Total PnL is typically driven by spread PnL (see tables \ref{tab:rl-exp-perf}, \ref{tab:rl-adaptive-perf}). The results illustrate that the RL agent is able to learn about its competitor's pricing strategy from the partial observations of its own pricing and resulting trades.

\begin{table}[b]
\caption{RL Agent vs Random/Persistent: Mean and standard deviation of $\epsilon_s$ and $\epsilon_b$ for RL agent}
\smallskip
\centering
\resizebox{.4\columnwidth}{!}{
\smallskip \begin{tabular}{lcc}
\hline 
  Competitor         &  $\epsilon_s$  &  $\epsilon_b$ \\
\hline
Persistent (-0.5)    &    -0.60 $\pm$ 0.16 &    -0.60 $\pm$    0.16 \\
Persistent (0.0)      &  -0.22 $\pm$ 0.15 &     -0.22 $\pm$  0.13 \\
Random ($\mathrm{Unif}[-0.5,0.5]$)   & -0.20 $\pm$   0.15 &  -0.20 $\pm$   0.15 \\
Random ($\mathrm{Unif}[-1.0,1.0]$)   & -0.03 $\pm$   0.14 &   -0.03 $\pm$   0.22 \\
\hline
\end{tabular}

}
\label{tab:rl-exp-eps}
\end{table}

\begin{figure}[t]
\centering
\begin{minipage}[t]{0.25\linewidth}
\includegraphics[width=\columnwidth]{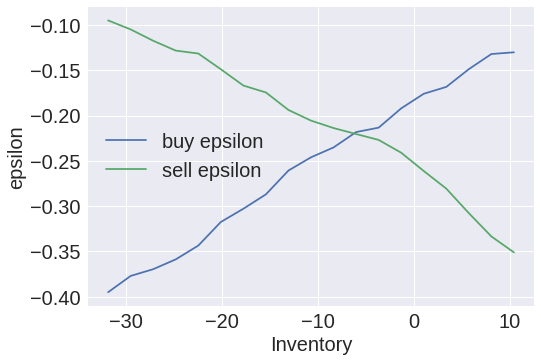}
\end{minipage}
\hfill
\begin{minipage}[t]{0.25\linewidth}
\includegraphics[width=\columnwidth]{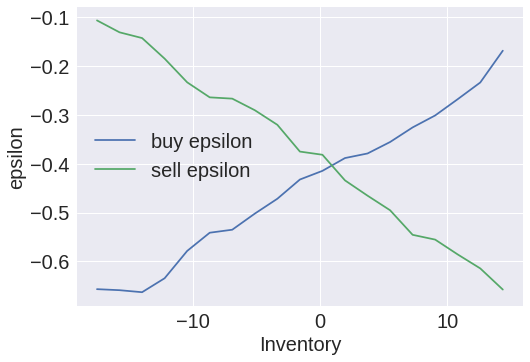}
\end{minipage}
\hfill
\begin{minipage}[t]{0.4\linewidth}
\includegraphics[width=\columnwidth]{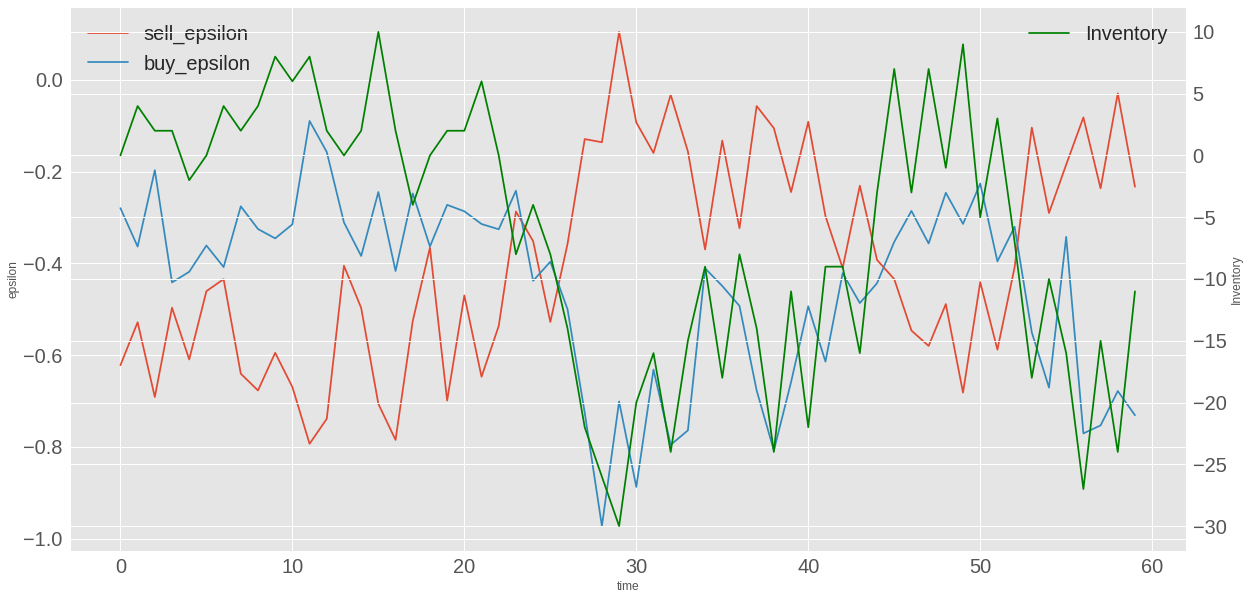} 
\end{minipage}
\caption{RL Agent - Price Skewing. Average $\epsilon_b$ (buy epsilon) and $\epsilon_s$ (sell epsilon) for RL agent at different levels of inventory. Left: vs persistent ($\epsilon=0$). Right: vs  adaptive ($\eta_{ms}=0.5$, $\gamma=2$).RL agent skewing: sample simulation against an adaptive agent}
\label{fig:rl-skew}
\end{figure}

\noindent \textbf{Price skewing for internalization}: The RL agent also learns to skew its pricing for internalization (see Figure \ref{fig:rl-skew}).  Unlike the adaptive agent that was explicitly designed to skew, the RL agent was given no prior knowledge or bias to enable skewing. Nevertheless, it learns to skew $\epsilon_b$ lower when inventory is negative to attract offsetting buys; similarly for positive inventory it skews $\epsilon_s$ lower to attract sells. 

\begin{figure}[t]
  \centering
  \begin{minipage}[t]{0.45\linewidth}
    \includegraphics[width=0.6\columnwidth]{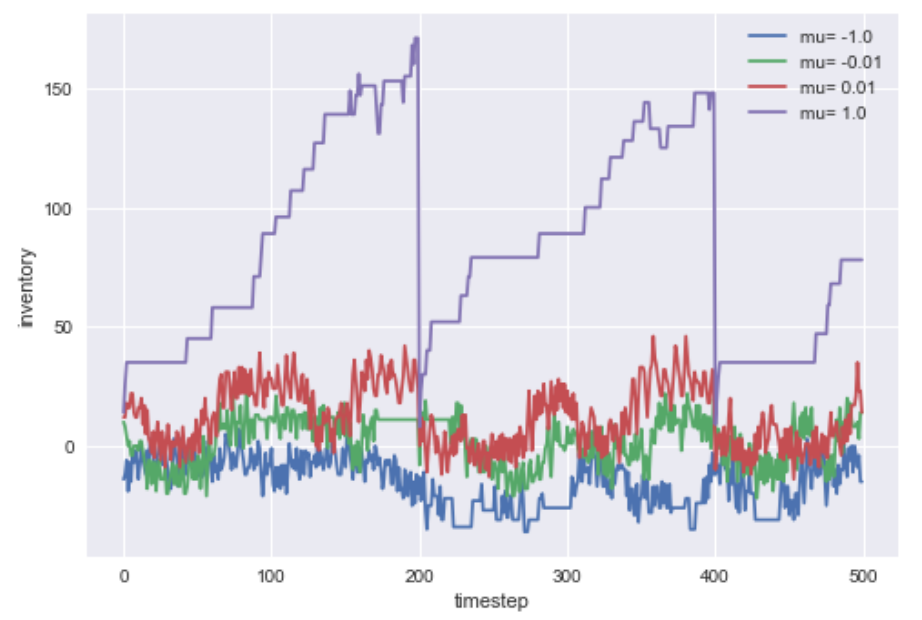} 
    \caption{RL Agent - inventory for various mid-price drift values $\mu$, as a function of time (500 timesteps, i.e. 2.5 rollouts of 200 timesteps)}
    \label{fig:rl-drift}
  \end{minipage}
  \hfill
  \begin{minipage}[t]{0.45\linewidth}
  \includegraphics[width=0.6\columnwidth]{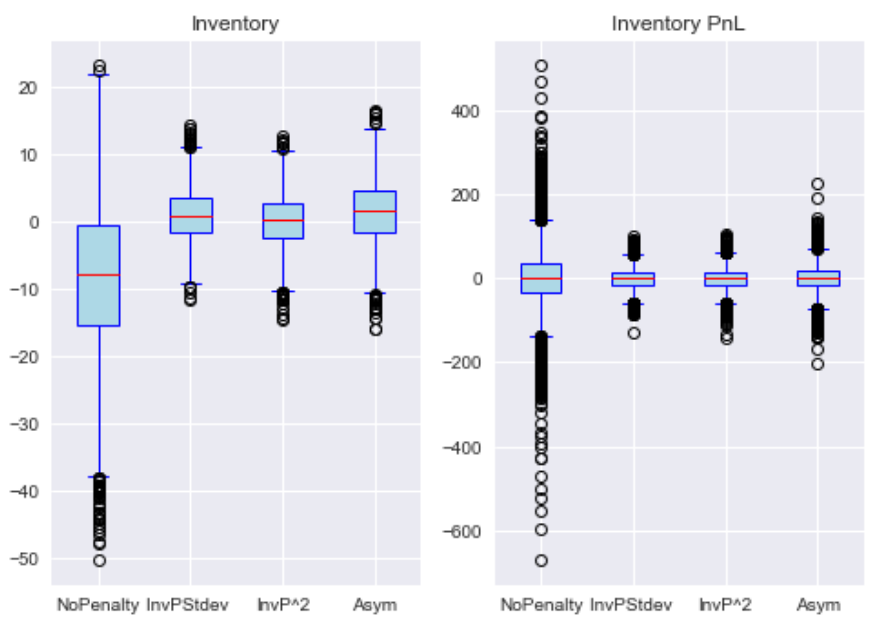} 
    \caption{RL Agent - Distributions of inventory and inventory PnL for different risk penalties}
    \label{fig:rl-rewardexp}
  \end{minipage}
\end{figure}

\subsection{Price environment}
\label{sec:rl-drift-exp}

The goal is to study the impact of the mid price drift $\mu$ on the inventory held by the RL agent (cf. section \ref{srefmidprice}). If the drift is positive (resp. negative), it is optimal to hold a positive (resp. negative) inventory as in average on every timestep, the mid-price will go up (resp. down), yielding a positive inventory PnL.
The setup consists of 3 agents: RL, Adaptive and persistent ($\epsilon_b=\epsilon_s=0$). 

The RL agent is able to learn that pattern well, in particular that i) the inventory held must be an increasing function of drift and ii) for extremely positive (resp. negative) values of the drift, the inventory held must be significantly positive (resp. negative). 

This is seen on figure \ref{fig:rl-drift} showing a zoom of the inventory over 500 rollout timesteps: for $\mu=1$, at each new rollout of 200 timesteps, the RL agent starts from 0 inventory and clearly tries to reach as quickly as possible the highest possible inventory, whereas when $\mu=-1$, the RL agent learned to hold a negative inventory (average of -16.06). For values of mu closer to 0, the inventory is closer to 0 as expected, and in average the inventory held for $\mu=0.01$ (10.82) is higher than in the case $\mu=-0.01$ (0.97).

\subsection{Reward function design}
\label{rl-reward-exp}

The goal is to study the impact of the penalty functions proposed in section \ref{sec:rl-agent} on the inventory held by the RL agent, as well as its inventory PnL. The setup consists of 2 agents: RL and persistent ($\epsilon_b=-0.5$, $\epsilon_s=0$), and the results are displayed in figure \ref{fig:rl-rewardexp}. 

It is observed that all 3 penalty functions have the desired effect of i) decreasing the inventory and inventory PnL standard deviation and ii) decreasing the absolute inventory mean value. 

\section{Conclusions and Future Work}

By using a multi-agent simulator to train an RL-based market maker agent with different competitive scenarios, reward formulations and market price drifts, we have been able to draw key insights into the behavior of the RL agent. We show that the reinforcement learning agent is able to learn about its competitor's pricing policy; it also learns to manage inventory by skewing prices and to maintain a positive (or negative) inventory depending on whether the market price drift is positive (or negative). We have also shown that different reward formulations can be used to create a risk averse RL-based market maker. This allows us to define a spectrum of market makers with different preferences and simulate more realistic competitive scenarios.

An interesting future direction of research, would be to study the effect of adverse selection and informed investors on market makers. Our multi-agent framework is particularly suitable for this because it explicitly models investor agents and allows market makers to counter adverse selection by tailoring their policies to the agent they are interacting with. 

\newpage
\bibliography{phantom}
\bibliographystyle{aaai}

 \section*{Disclaimer}
 This paper was prepared for information purposes by the AI Research Group of JPMorgan Chase \& Co and its affiliates (“J.P. Morgan”), and is not a product of the Research Department of J.P. Morgan.  J.P. Morgan makes no explicit or implied representation and warranty and accepts no liability, for the completeness, accuracy or reliability of information, or the legal, compliance, financial, tax or accounting effects of matters contained herein.  This document is not intended as investment research or investment advice, or a recommendation, offer or solicitation for the purchase or sale of any security, financial instrument, financial product or service, or to be used in any way for evaluating the merits of participating in any transaction.   

\section*{Appendix}
\subsection{Proof of Theorem \ref{thmopteps}}
\label{proofthm}
Since the spread PnL is the sum of the sell side and buy side spread PnL, and since we can act independently on the two sides via $\epsilon_b$, $\epsilon_s$, it is maximized when both are maximized. Let us then prove the theorem for the sell side, the proof for the buy side is identical. Denote $1_{\{\cdot\}}$ the indicator function and $v_i$, $i=1..N$ the N investor trades (possibly random but independent from $\epsilon_s$). Then, there are 3 possibilities: either $\epsilon_s>\epsilon^{comp}_s$ in which case the RL agent gets 0 trades; $\epsilon_s<\epsilon^{comp}_s$ in which case the RL agent gets all the trades, and $\epsilon_s=\epsilon^{comp}_s$ in which case for each trade, there is a probability $p$ to get it: we model the latter by Bernoulli(p) random variables $y_i$, $i=1..N$, independent from all other random variables, such that the investor trades in case of $\epsilon_s=\epsilon^{comp}_s$ are given by $y_i v_i$. We get:
$$
\mbox{SpreadPnL}(\epsilon_s)=\left(\sum_{i=1}^N |v_i| S_{ref}(v_i)(1+\epsilon_s)\right)1_{\{\epsilon_s<\epsilon^{comp}_s\}} 
$$
$$
+ \left(\sum_{i=1}^N y_i |v_i| S_{ref}(v_i)(1+\epsilon_s)\right)1_{\{\epsilon_s=\epsilon^{comp}_s\}}
$$
Taking expectations, using independence between $\epsilon^{comp}_s$, $y_i$, $v_i$, and denoting $\alpha_N:=\sum_{i=1}^N E[|v_i| S_{ref}(v_i)]$ we get:
$$
E[\mbox{SpreadPnL}(\epsilon_s)]=(1+\epsilon_s) \alpha_N P[\epsilon_s<\epsilon^{comp}_s]
$$
$$
+ (1+\epsilon_s)\alpha_N E[y_i] P[\epsilon_s=\epsilon^{comp}_s]
$$
Since $E[y_i]=p$, $P[\epsilon_s<\epsilon^{comp}_s]=1-F^{comp}_{s}(\epsilon_s)$, and $P[\epsilon_s=\epsilon^{comp}_s]=F^{comp}_{s}(\epsilon_s)-\lim_{x \to \epsilon_s^-} F^{comp}_{s}(x)$ (where as usual $x \to \epsilon_s^-$ denotes the left limit), we finally get:
$$
\epsilon^*_s= \mbox{argmax}_\epsilon h(\epsilon): (1+\epsilon) (1-F^{comp}_{s}(\epsilon)
+p \hat{F}^{comp}_{s}(\epsilon))
$$
In particular if $F^{comp}_{s}$ is continuous, we get $\hat{F}^{comp}_{s}=0$. 

In the case $\mathrm{Unif}[a, b]$, we have $\hat{F}^{comp}_{s}=0$ and  $F^{comp}_{s}(\epsilon)=\frac{\epsilon-a}{b-a}$ for $\epsilon \in [a,b]$. The maximum of the function $h$ on $[-1, \infty)$ is the maximum of that function on $[a,b)$, since for $\epsilon<a$, $h(\epsilon)=1+\epsilon \leq h(a)=1+a$ and for $\epsilon \geq b$, $h(\epsilon)=0$. Standard calculus shows that the function $\epsilon \to (1+\epsilon)\frac{b-\epsilon}{b-a}$ has a global maximum located at $\epsilon^*=\frac{b-1}{2}$. Hence, the maximum of $h$ on $[a,b)$ is $\epsilon^*$ if $\epsilon^* \geq a$, and $a$ otherwise (note that we always have $\epsilon^* \leq b$ since $b \geq -1$).

\subsection{Simulation of the exchange reference price}
\label{sref}
In this section we use the terminology "bid" (resp. "ask") for "buy" (resp. "sell").

\begin{figure}[t]
  \centering
  \begin{minipage}[t]{0.45\linewidth}
    \includegraphics[width=0.6\columnwidth]{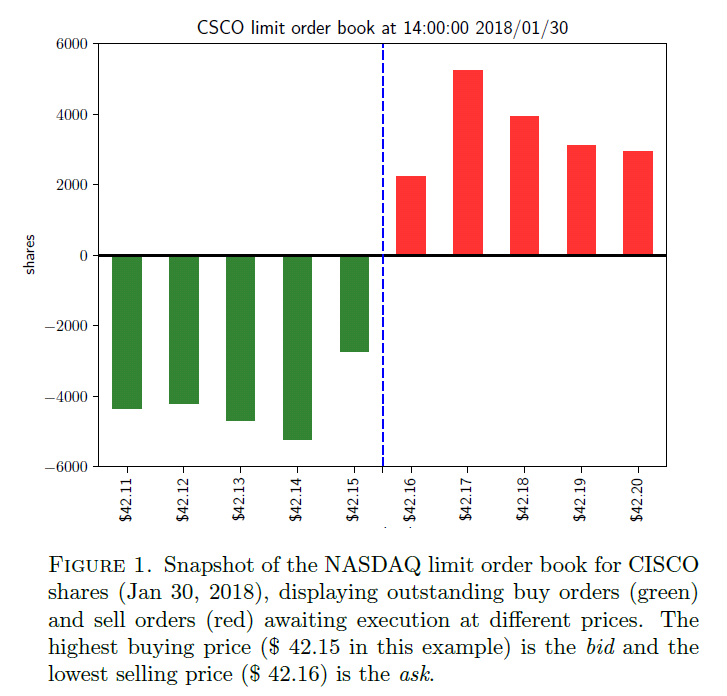} 
    \caption{Illustration of a LOB as displayed in \cite{CM}}
    \label{lob}
  \end{minipage}
  \hfill
  \begin{minipage}[t]{0.45\linewidth}
  \includegraphics[width=0.6\columnwidth]{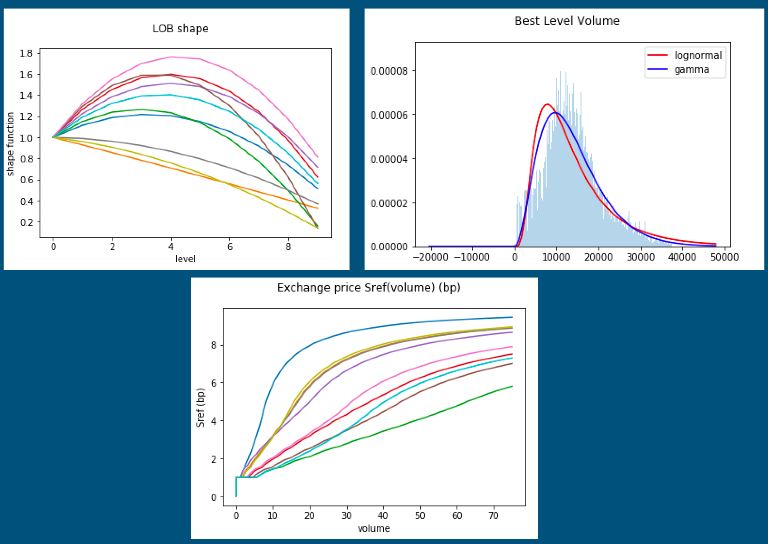} 
    \caption{calibration of $S_{ref}$}
    \label{srefg}
  \end{minipage}
\end{figure}

The market reference price curve $S_{ref}(v)$ is assumed to be known by all participants and specifies the cost of trading a size $v$ on the reference exchange (relatively to the mid price), which operates like a limit order book (cf. figure \ref{lob}). Such a structure is characterized by 2 lists of 2-tuples $(n^a_k, a_k)$, $(n^b_k, b_k)$, $k \geq 0$, where:
\begin{itemize}
    \item (orders available on the ask side) one can buy at most a size of $n^a_k$ at price level $a_k$, where $(a_k)$ is an increasing sequence and $a_0$ is called the ask price, i.e. the lowest price at which you can buy (but only at most a size of $n^a_0$).
    \item (orders available on the bid side) one can sell at most a size of $n^b_k$ at price level $b_k$, where $(b_k)$ is a decreasing sequence and $b_0$ is called the bid price, i.e. the highest price at which you can sell (but only at most a size of $n^b_0$).
\end{itemize}

For example, if you want to buy a total size of $v$, the related cost will be given by:
$$
(v-\sum_{k=0}^{n^*-1}n^a_k)a_{n^*}+\sum_{k=0}^{n^*-1}a_k n^a_k
$$
$$
n^*:=inf\{n: v<\sum_{k=0}^{n}n^a_k\}
$$
The reference market spread is defined as $s_0:=a_0-b_0=2S_{ref}(0)$, and the reference market mid price is given by $P_t=\frac{a_0+b_0}{2}$. In the following we will assume that $S_{ref}(v)$ is the same on the bid and ask side, so we continue the reasoning with the ask side. 

The reference market spread $s_0$ is simulated, at each time step, as a normal random variable of mean $0.01\%$ and stdev $0.005\%$ (we further enforce $s_0$ to be no less than $0.001\%$ and no more than $0.025\%$). 

We model 10 levels of the bid (resp. ask ) side, given by $\frac{s_0}{2}+n$ b.p. (1 b.p.=0.01\%), $n=0..9$. The volume $v(n)$ available at level $n$ is modeled as:
$$
v(n)=\underbrace{\lambda}_{\mbox{scaler}} \cdot \underbrace{v(0)}_{\mbox{best level volume}}\cdot \underbrace{shape(n)}_{\mbox{shape function}}
$$
$$
shape(0)=1, \hspace{2mm} v(0) \sim \mbox{Gamma Law}, \hspace{2mm} E[v(0)]=1.
$$

The scaler $\lambda$ is a constant currently calibrated before the experiment such that investor trades process are able to hit the full range of LOB levels. $v(0)$ and the shape function are random variables calibrated to LOB data (1 day of stock MSFT). We smooth the raw LOB shape data by fitting a 2nd order polynomial to it, and we sample independently the shape and v(0) at each timestep. This methodology is illustrated in figure \ref{srefg}, where various sample shapes and generated $S_{ref}$ prices are presented, together wit a Gamma fit of $v(0)$.


\subsection{RL agent reward convergence}
\subsubsection{Random Agent Competitor}
\begin{figure}[t]
  \centering
  \begin{minipage}[t]{0.45\linewidth}
    \includegraphics[width=0.6\columnwidth]{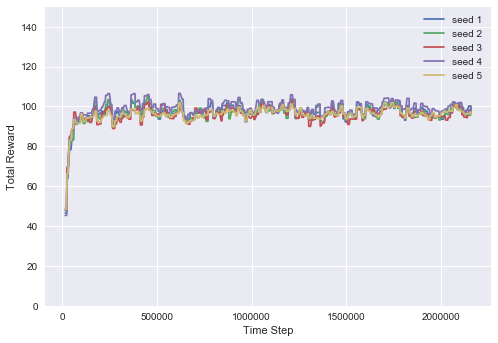} 
    \caption{RL Agent - trained against Random Agent(-1,1) with 5 different seeds}
    \label{fig:seedrandom1}
  \end{minipage}
  \hfill
  \begin{minipage}[t]{0.45\linewidth}
  \includegraphics[width=0.6\columnwidth]{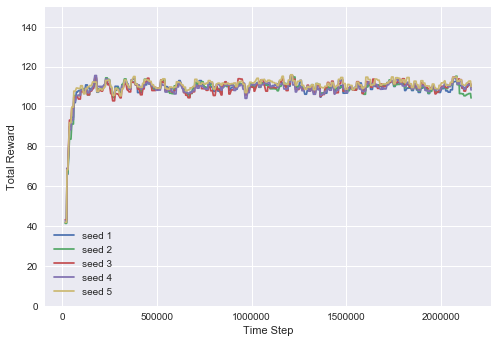} 
    \caption{RL Agent - trained against Random Agent(-0.5,0.5) with 5 different seeds}
    \label{fig:seedrandom2}
  \end{minipage}
\end{figure}

The goal of this section is to show the convergence of the RL reward function in the learning phase with respect to the number of timesteps, and so for 5 different random number generator seeds. We display in figures \ref{fig:seedrandom1} (resp. \ref{fig:seedrandom2}) the reward functions of the RL agent in the learning phase in the case where it competes against a random agent quoting random prices $\epsilon_s$, $\epsilon_b$ between -1 and 1 (resp. -0.5 and 0.5). The reward function is shown to converge.

\subsubsection{Persistent Agent Competitor}
\begin{figure}[t]
  \centering
  \begin{minipage}[t]{0.45\linewidth}
    \includegraphics[width=0.6\columnwidth]{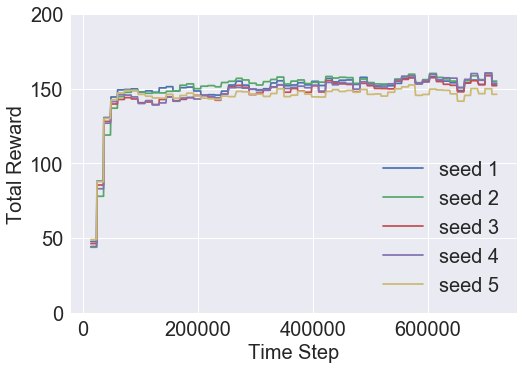} 
    \caption{RL Agent - trained against Persistent Agent $\epsilon_s=\epsilon_b=0$ with 5 different seeds}
    \label{fig:seedp1}
  \end{minipage}
  \hfill
  \begin{minipage}[t]{0.45\linewidth}
  \includegraphics[width=0.6\columnwidth]{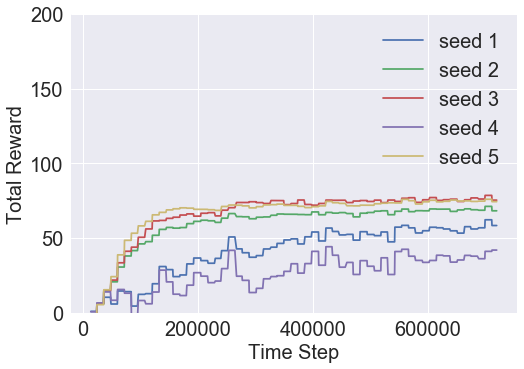} 
    \caption{RL Agent - trained against Persistent Agent $\epsilon_s=\epsilon_b=-0.5$ with 5 different seeds}
    \label{fig:seedp2}
  \end{minipage}
\end{figure}

The goal of this section is to show the convergence of the RL reward function in the learning phase with respect to the number of timesteps, and so for 5 different random number generator seeds. We display in figures \ref{fig:seedp1} (resp. \ref{fig:seedp2}) the reward functions of the RL agent in the learning phase in the case where it competes against a persistent agent quoting prices $\epsilon_s=\epsilon_b=0$ (resp. $\epsilon_s=\epsilon_b=-0.5$). The reward function is shown to converge.

\subsection{RL agent hedging}

In order to illustrate the ability of the RL agent to outperform its adaptive counterpart in specific situations, we study the case where hedge cost is zero (i.e. it doesn't cost anything to hedge inventory), $S_{ref}$ is high so that one is incentivized to trade with investors, but on the other hand, investors are toxic in the sense that they have perfect knowledge of the mid price move on future timesteps, meaning that inventory PnL will always be negative. This setting illustrates the case where e.g. hedge funds possess complex statistical models that allow them to predict accurately future price moves. The adaptive agent is set to 50\% target market share and risk aversion of 2.

\begin{figure}[t]
\centering
\begin{minipage}[t]{0.5\linewidth}
\includegraphics[width=\columnwidth]{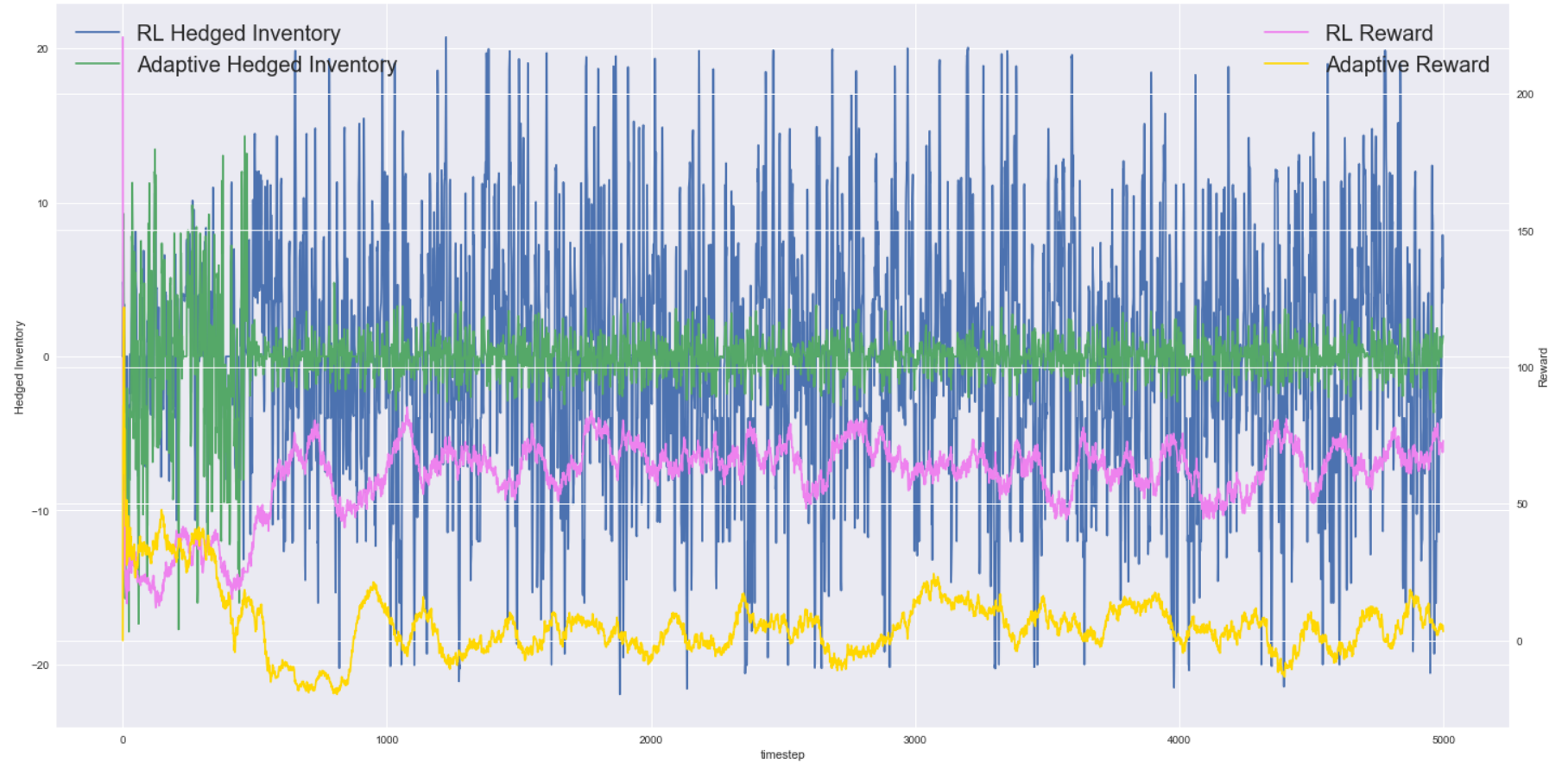}
\end{minipage}
\hfill
\caption{RL Agent Vs. Adaptive agent - toxic investors -  hedged inventory and reward.}
\label{fig:rl-hedgebehavior}
\end{figure}

The adaptive agent currently doesn't keep track of toxicity (i.e. the ability of an investor to predict mid-price move), so it is expected of the RL agent to exploit that weakness. We display in figure \ref{fig:rl-hedgebehavior} the hedged inventory of both agents as a function of timesteps, as well as the total rewards. It is seen that the RL agent indeed learns to hedge significantly more than the adaptive agent in this setting, so that it is able to outperform it.

\subsection{Adaptive agent skewing}

\begin{figure}[t]
  \centering
  \begin{minipage}[t]{0.45\linewidth}
    \includegraphics[width=0.6\columnwidth]{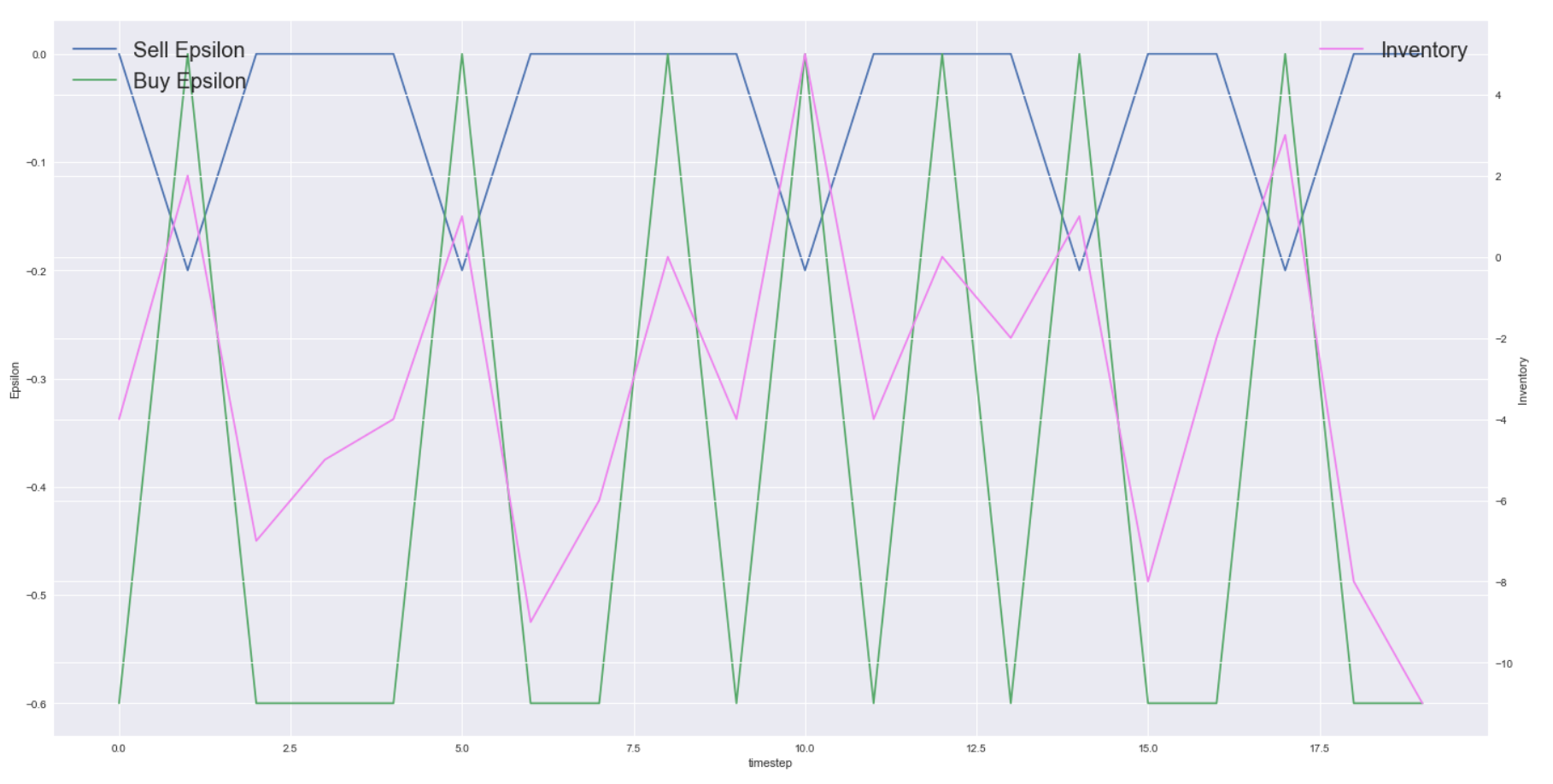} 
    \caption{Adaptive agent - $\epsilon_b$, $\epsilon_s$ and Inventory - Risk Aversion=0}
    \label{fig:adskew0}
  \end{minipage}
  \hfill
  \begin{minipage}[t]{0.45\linewidth}
  \includegraphics[width=0.6\columnwidth]{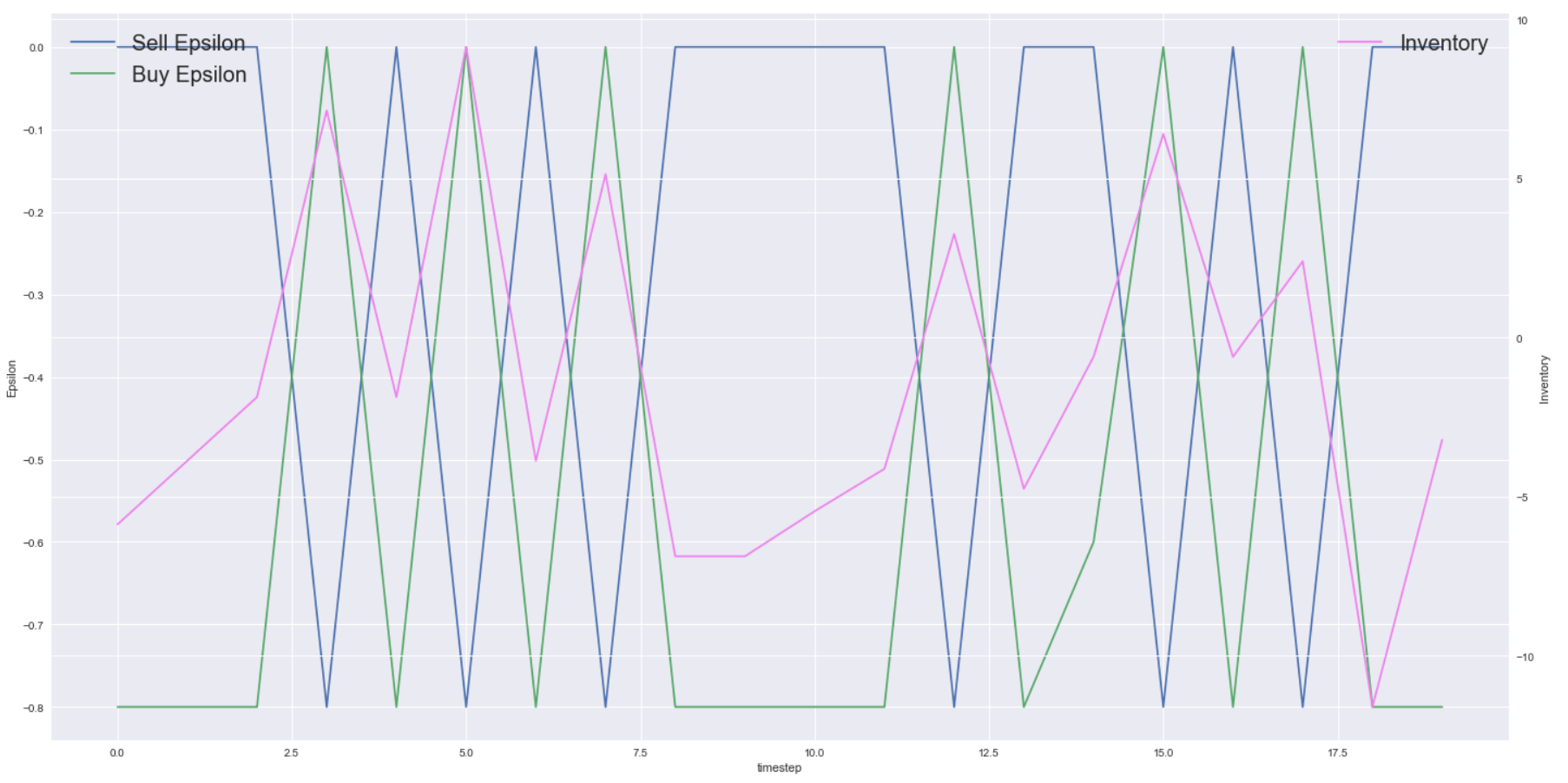} 
    \caption{Adaptive agent - $\epsilon_b$, $\epsilon_s$ and Inventory - Risk Aversion=2}
    \label{fig:adskew}
  \end{minipage}
\end{figure}

We illustrate in figures \ref{fig:adskew0}, \ref{fig:adskew} the pricing mechanism of the adaptive agent for risk aversions 0, 2. In this setting, the adaptive agent has a target market share of 25\% and competes against a persistent agent quoting $\epsilon_s=0$, $\epsilon_b=-0.5$. When the adaptive agent quotes $\epsilon_s=\epsilon_b=0$, he would then in average achieve 25\% market share (remember that when 2 agents quote the same price, they have probability 50\% to be selected by the investor). Hence, in the adaptive pricing policy, if there was no step 2 but only step 1 targeting market share, $\epsilon_s=\epsilon_b=0$ would be the optimal level, and this is why in figures \ref{fig:adskew0}, \ref{fig:adskew}, $\max(\epsilon_b,\epsilon_s)=0$, i.e. at every timestep, the maximum of the blue and green lines correspond to step 1 of the adaptive pricing policy. 

The reason why $\epsilon_b-\epsilon_s \neq 0$ (i.e. skewing) is because of step 2 which consists of optimizing a trade-off between spread PnL and aversion to inventory PnL. When risk aversion is zero (figure \ref{fig:adskew0}), $\min(\epsilon_b,\epsilon_s)$ is -0.6 when $\min(\epsilon_b,\epsilon_s)=\epsilon_b$, and -0.2 when $\min(\epsilon_b,\epsilon_s)=\epsilon_s$: that is, when risk aversion is zero, the market maker only tries to optimize spread PnL and so it suffices to price just slightly better than its competitor (which prices at $\epsilon_s=0$, $\epsilon_b=-0.5$). When risk aversion is 2 (figure \ref{fig:adskew}), the market maker is eager to lower its absolute inventory at the expense of its spread PnL, and that is why in that case, the market maker prices aggressively in order to attract investor flow: in this case we always have $\min(\epsilon_b,\epsilon_s)=-0.8$ so that the adaptive market maker is guaranteed to attract a significant amount of investor flow in order to reduce its absolute inventory.

\subsection{Adaptive agent hedging policy}

Table \ref{tab:adaptive-hedging} demonstrates that our adaptive hedging policy outperforms alternative policies - persistent, random and a similar adaptive policy which ignores internalization (i.e. $v_{\epsilon}=0$). The test was conducted by evaluating the performance of an agent with an adaptive pricing policy but different hedging policies, against a persistent agent ($\epsilon_b=\epsilon_s=0$, $x=0.2$). As expected, ignoring the effect of internalization leads to over-hedging and poor performance especially when the risk aversion is high (hedge cost of -43 for $\gamma=2)$. 

\subsection{Adaptive vs Random/Persistent Agents}
\label{sec-adaptive-exp}

We validated the effectiveness of the adaptive agent by evaluating it against a random ( $\epsilon_b = \epsilon_s \sim \mathrm{Unif}[-1, 1]$) and persistent ($\epsilon_b= -0.5$, $\epsilon_s= 0 $, $x=0.2$) agent. In each case, we tested different adaptive policies, specified by the market share target and risk aversion level for the adaptive agent. 

   \begin{table}[t]
         \begin{minipage}{0.45\linewidth}
         \caption{Adaptive vs. Persistent Agent: Average excess PnL per time-step of adaptive agent over competitor}
         \label{tab:adaptive-exp-1}
          \centering
             \resizebox{\columnwidth}{!}{\smallskip \begin{tabular}{llrrrr}
\hline 
MS target & Risk Aversion & Spread PnL & Inventory PnL & Hedge Cost & Total PnL\\
\hline 
0.25      & 0             & 14.64      & 0.02      & 8.67       & 23.33  \\
          & 1             & 10.34      & -0.06     & 8.21       & 18.48  \\
          & 2             & 0.61       & 0.03      & 5.19       & 5.83   \\
0.50       & 0             & 55.7       & 0.14      & 2.77       & 58.61  \\
          & 1             & 47.75      & -0.09     & 2.58       & 50.23  \\
          & 2             & 34.01      & -0.15     & -2.18      & 31.68  \\
\hline
\end{tabular} }
         \end{minipage}
         \hfill
         \begin{minipage}{0.45\linewidth}
         \centering
         \caption{Adaptive vs. Random Agent: Average excess PnL of per time-step of adaptive agent over competitor}
         \label{tab:adaptive-exp-2}
             \resizebox{\columnwidth}{!}{\smallskip \begin{tabular}{llrrrr}
\hline 
MS target & Risk Aversion & Spread PnL & Inventory PnL & Hedge Cost & Total PnL\\
\hline 
0.25      & 0             & 13.85      & -0.03     & 28.98      & 42.81  \\
          & 1             & -7.3       & -0.1      & 28.32      & 20.92  \\
          & 2             & -21.35     & -0.07     & 25.66      & 4.24   \\
0.50      & 0             & 28.81      & 0.25      & 25.38      & 54.44  \\
          & 1             & 20.84      & 0.02      & 24.88      & 45.74  \\
          & 2             & 17.05      & -0.03     & 17.19      & 34.22  \\
\hline 
\end{tabular}}
         \end{minipage}
         \hfill
         \vspace{5mm}
         \begin{minipage}{0.45\linewidth}
         \centering
         \caption{Adaptive Hedging Policy vs Alternatives: Average PnL per time-step (multiplied by 100).}
         \label{tab:adaptive-hedging}
             \resizebox{\columnwidth}{!}{\smallskip \begin{tabular}{llrrrr}
\hline 
Hedging Policy    & Spread PnL & Inventory & Hedge Cost & Total PnL  \\
  & & PnL &  & \\
\hline 
Random: $x \sim [0,1]$        & 51.36      & 0.07      & -15.70     & 35.73  \\
Persistent: $x = 0.2$         & 51.43      & -0.03     & -3.84      & 47.57  \\
Adaptive: $\gamma=1$          & 51.39      & 0.02      & -0.22      & 51.19  \\
Adaptive: $\gamma=2$          & 51.34      & 0.10      & -1.23      & 50.21  \\
Adaptive $v_{\epsilon}=0$: $\gamma=1$        & 51.53      & -0.07     & -2.30      & 49.17  \\
Adaptive: $v_{\epsilon}=0$: $\gamma=2$      & 50.56      & -0.01     & -43.47     & 7.08   \\
\hline 
\end{tabular}}
         \end{minipage}
     \end{table}

 As shown in tables \ref{tab:adaptive-exp-1} and \ref{tab:adaptive-exp-2}, the adaptive agent is able to outperform its competitors in terms of Total PnL; in most cases it has an advantage on both spread PnL and hedge cost.  However, increasing the risk aversion decreases the competitiveness of the adaptive agent because it tends to hedge more and adjust pricing to maximize internalization rather than spread PnL (in Step 2 outlined in \ref{sec:adaptive-pricing-policy}). 

\end{document}